# Designing Efficient Metal Contacts to Two-Dimensional Semiconductors MoSi₂N₄ and WSi₂N₄ Monolayers

Qianqian Wang[1,2†], Liemao Cao[3†‡], Shi-Jun Liang[4], Weikang Wu[2,5], Guangzhao Wang[1,2], Ching Hua Lee[6], Wee Liat Ong[7,8], Hui Ying Yang[9], Lay Kee Ang[1], Shengyuan A. Yang[1,2‡], Yee Sin Ang[1*]

1.  Science, Mathematics and Technology, Singapore University of Technology and Design, Singapore 487372, Singapore.

2.  Research Laboratory of Quantum Materials, Singapore University of Technology and Design, Singapore 487372, Singapore.

3.  College of Physics and Electronic Engineering, Hengyang Normal University, Hengyang 421002, China.

4.  National Laboratory of Solid State Microstructures, School of Physics, Collaborative Innovation Center of Advanced Microstructures, Nanjing University, Nanjing, 210093, China

5.  Division of Physics and Applied Physics, School of Physical and Mathematical Sciences, Nanyang Technological University, Singapore, 637371, Singapore

6.  Department of Physics, National University of Singapore, Singapore 117542, Singapore.

7.  Zhejiang University-University of Illinois at Urbana-Champaign Institute (ZJU-UIUC) College of Energy Engineering, Zhejiang University, Hangzhou, Zhejiang 310027, China.

8.  State Key Laboratory of Clean Energy Utilization, Zhejiang University, Hangzhou, Zhejiang, 310027, China.

9.  Engineering Product Development, Singapore University of Technology and Design, Singapore 487372, Singapore.

**Abstract.** Metal contacts to two-dimensional (2D) semiconductors are ubiquitous in modern electronic and optoelectronic devices. Such contacts are, however, often plagued by strong Fermi level pinning (FLP) effect which reduces the tunability of the Schottky barrier height (SBH) and degrades the performance of 2D-semiconductor-based devices. In this work, we show that monolayer $MoSi_2N_4$ and $WSi_2N_4$ – a recently synthesized 2D material class with exceptional mechanical and electronic properties – exhibit strongly suppressed FLP and wide-range tunable SBH when contacted by metals. An exceptionally large SBH slope parameter of $S$=0.7 is obtained, which outperform the vast majority of other 2D semiconductors. Such surprising behavior arises from the unique morphology of $MoSi_2N_4$ and $WSi_2N_4$. The outlying Si-N layer forms a native atomic layer that protects the semiconducting inner-core from the perturbance of metal contacts, thus suppressing the FLP. Our findings reveal the potential of $MoSi_2N_4$ and $WSi_2N_4$ monolayers as a novel 2D material platform for designing high-performance and energy-efficient 2D nanodevices.

†These authors contributed equally

‡Corresponding Author. liemao_cao@hynu.edu.cn

§Corresponding Author. shengyuan_yang@sutd.edu.sg

*Corresponding Author. yeesin_ang@sutd.edu.sg

**Introductions.** Electrical contacts between metals and semiconductors are ubiquitous in modern electronic and optoelectronic devices. An interfacial potential barrier, known as the Schottky barrier (SB), is commonly formed at the metal/semiconductor interface. In electronics and optoelectronics applications, the presence of a sizable SB, typically larger than a few $k_B T$, can severely impede the charge injection efficiency [1]. A SB is intimately linked to the contact resistance at the metal/semiconductor contact [2]. In the thermionic charge injection regime [3,4], the contact resistance is exponentially raised by the Schottky barrier height (SBH), i.e. $R_c^{(TE)} \propto$ exp $(\Phi_B / k_B T)$ where $\Phi_B$ is the SBH [1,5]. An overly large $R_c$ raises the power dissipation at the contact, lowers the 'on' state current, $I_{on}$, and increases the device delay time as well as the dynamical power consumption [6]. Reducing the SBH at the metal/semiconductor contact has thus become one of the key challenges towards energy-efficient and high-speed semiconductor devices [7].

In a metal/semiconductor contact, SBH arises from the mismatch between the metal work function ($W_M$) and the semiconductor electron affinity $E_{ea}$ ($n$-type Schottky contact) or ionization potential $E_{ip}$ ($p$-type Schottky contact). Taking into account the inevitable presence of metal/semiconductor interactions and interfacial defects, the SBH across a metal/semiconductor contact can be phenomenologically captured by the modified Schottky-Mott (SM) rule [3,8–10],

$$\Phi_{Be} = S_e(W_M - E_{ea}) + c_e \qquad (1)$$

$$\Phi_{Bh} = S(E_{ip} - W_M) + c_h \qquad (2)$$

where the subscript '$e$' and '$h$' denote $n$-and $p$-type contacts, respectively, $c_{e(h)}$ is a material-and contact-dependent term [8], and $\Phi_{Be}$ and $\Phi_{Bh}$ is the electron-type and hole-type SBH, respectively. Here, $S_{e(h)} \equiv |d\Phi_{Be(h)}/dW_M|$ is the *slope parameter* – an important phenomenological parameter widely used in characterizing the deviation of the SBH from the ideal SM limit ($S = 1$). In realistic metal/semiconductor contacts, $S \ll 1$ due to the presence of multiple nonideal factors, such as the formation of metal-induced gap states, defect-induced gap states, mid-gap states and interface dipole, as well as the modifications of the electronic band structures of the semiconductor when contacted by metals [8,10,11]. In this case, the SBH is pinned to a narrow range of value – an adverse effect commonly known as the *Fermi level pinning* (FLP). Strong FLP poses a great challenge to the design of semiconductor electronics and optoelectronics. Suppressing FLP has thus become a centerpiece in the design of high-performance semiconductor devices.

In the few-atom-thick limit, two-dimensional (2D) semiconductors, such as $MoS_2$ and $WS_2$ [12], continue to be plagued by FLP [17,18]. Although 2D semiconductors and their van der Waals heterostructures has shown great promises in low-power electronics [13], optoelectronics [14] and neuromorphic applications [15,16], the lack of wide-range tunable SBH – commonplace in most 2D semiconductors – has severely impeded the development of high-performance nanodevices. Despite tremendous theoretical and experimental efforts, designing high-efficiency electrical contacts to 2D semiconductor with wide-range tunable SBH remains an ongoing challenge. Particularly, the metallization of a 2D semiconductor by the contacting metal often substantially alters the electronic structures of the heterostructure via the generation of mid-gap states [19,20], causing strong FLP effect that leads to a poor SBH tunability [20]. For the vast majority of 2D semiconductors, $S$ is typically less than 0.4 as predicted by density functional theory (DFT) calculations [19–33], and are even lower in experimental measurements due to the inevitable presence of defects at the contact interface [34–36]. To resurrect a wide-range tunable

SBH in 2D semiconductors, atomically sharp van der Waals (VDW) metal/2D-semiconductor electrical contacts have been proposed [37]. Such *VDW-type contacts* harness the weak VDW interfacial coupling to reduce the metal/semiconductor interactions, yielding an $S \approx 1$ that approaches the ideal SM limit [38–40]. However, as VDW-type contact often involves complex fabrication techniques, a 2D semiconductor class that is *inherently* achieve Fermi level *unpinning* without necessarily relying on the VDW contact paradigm remains elusive thus far.

In this work, we perform a first-principle density functional theory (DFT) investigation [41–47] and show that, in contrary to the common knowledge that 2D semiconductors are prone to strong FLP, the recently discovered *synthetic* 2D monolayers, $MoSi_2N_4$ and $WSi_2N_4$ [48,49], exhibit strongly suppressed FLP and excellent SBH tunability without relying on the VDW-type contact engineering. The SBH is widely tunable in $MoSi_2N_4$ and $WSi_2N_4$ monolayers, reaching an exceptionally high slope parameter of $S = 0.69$ and $S = 0.77$, respectively – a value much larger than other commonly studied 2D semiconductors. The FLP suppression originates from the unique morphology of $MoSi_2N_4$ and $WSi_2N_4$ monolayers, in which the semiconducting states residing in the Mo-S or W-S inner core-layer are protected by the outlying Si-N atomic layers – an intriguing FLP suppression mechanism not found in other classes of 2D semiconductors. Our results reveal $MoSi_2N_4$ and $WSi_2N_4$ monolayer as an unusual 2D semiconductor class with built-in atomic layer protection, thus opening up an alternative and complementary route to the VDW contact paradigm towards efficient SBH tuning and high-performance electrical contact engineering.

**Electronic properties and the intrinsic atomic layer protection of the semiconducting states.**

The 2D monolayers of $MoSi_2N_4$ and $WSi_2N_4$ [Figs. 1(a) and 1(b)] belong to the family of $MA_2X_4$

synthetic 2D layered materials with no known 3D parent materials (M = early transition metal, A = Si or Ge, X = N, P or As). The $MA_2X_4$ 2D material family covers a wide variety of semiconducting, metallic, insulating and magnetic phases, thus offering an exciting new platform for the exploration 2D semiconductor physics and device applications [50–54]. $MoSi_2N_4$ and $WSi_2N_4$ monolayers are indirect band gap semiconductor with band gap values of 1.73 eV and 2.06 eV, respectively [Figs. 1(c) and 1(d)], and exhibit excellent structural stability and mechanical strength [48].

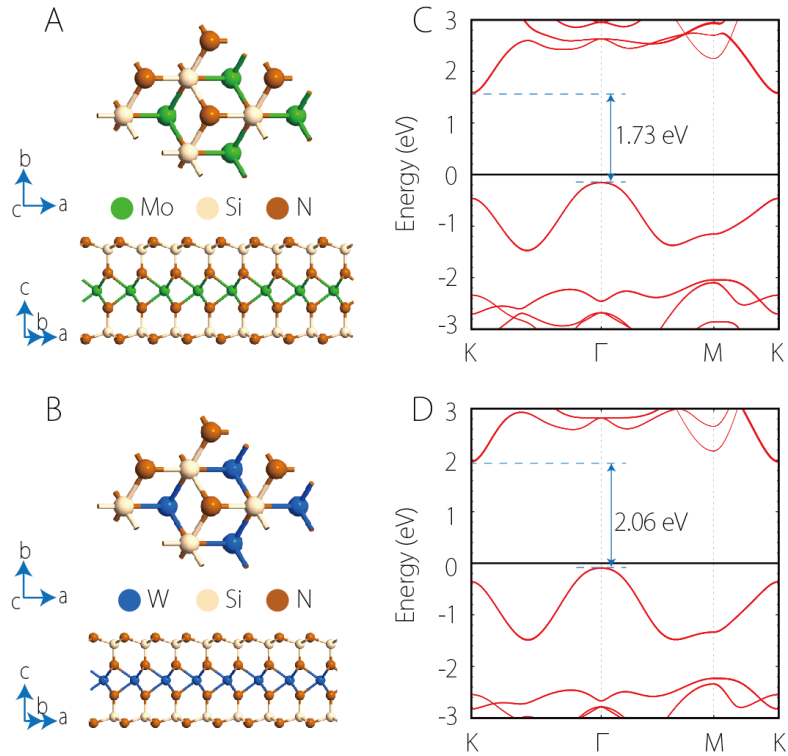

**Figure 1.** Crystal and electronic band structures of isolated $MoSi_2N_4$ and $WSi_2N_4$ monolayers, and a comparison of the slope parameter (*S*) for various 2D semiconductors. Top and side view of (A) $MoSi_2N_4$ monolayer and (B) $WSi_2N_4$ monolayer. Electronic band structures of isolated (C) $MoSi_2N_4$ and (D) $WSi_2N_4$ monolayers.

The $MoSi_2N_4$ ($WSi_2N_4$) monolayer is composed of a Mo-S (W-S) monolayer sandwiched by two Si-N layers. We perform first-principle DFT simulations to model the contact heterostructure composed of $MoSi_2N_4$ and $WSi_2N_4$ monolayers contacted by a large variety of metals (3D metals Sc, In, Ti, Ag, Cu, Ni, Au, Pd, Pt; 2D semimetal graphene; and 2D metal $NbS_2$ monolayer) with work function ranging from 3.3 eV to 6.0 eV (see *Supporting Information* for the DFT simulation methods). We found that metal contacts to $MoSi_2N_4$ and $WSi_2N_4$ cover a large variety of contact types, including *n*-type Schottky contact, *p*-type Schottky contact, and Ohmic contact with zero SBH. The interlayer distance between the metal and 2D monolayers are less than 3 Å for Cu, Pd, Ti and Ni (see Table S1), suggesting the prevalence of non-VDW-type metal contacts. We take Au and Ti metal contacts as illustrative examples of Schottky (Fig. 2) and Ohmic contacts (Fig. 3), respectively (see Fig. S1 for the relaxed band structures of all metal contacts calculated in this work). The lattice structures of the Au and Ti contacts to $MoSi_2N_4$ and $WSi_2N_4$ monolayers are shown in Figs. 2(a) and 3(a), respectively.

A closer inspection on the electronic band structures of the metal/$MoSi_2N_4$ and metal/$WSi_2N_4$ reveals intriguing behaviors. The metal/semiconductor heterostructures is nearly free of mid-gap states for all metals as revealed in the projected band structures and the partial density of states (PDOS) [Figs. 2(b) and 3(e) for Au Schottky contacts, and Figs. 3(b) and 3(e) for Ti Ohmic contact, see also Fig. S2 for the projected band structures and PDOS of other metal contacts]. By projecting the 2D semiconductor electronic states onto the band structures, we find that the semiconducting bands, particularly the electronic states around the valence band maximum (VBM) and the conduction band minimum (CBM), remain intact in all metal contacts, suggesting the near absence of mid-gap states within the semiconductor band gap. For Ti contact, despite being a close contact type with exceedingly small interlayer distance of 1.75 Å and 1.78

Å, mid-gap states remain nearly absent [see PDOS in Fig. 3(b) and 3(e)], which is in stark contrast to metal-contacted $MoS_2$ where large abundance of mid-gap states localizing in the Mo sites are created [20,55].

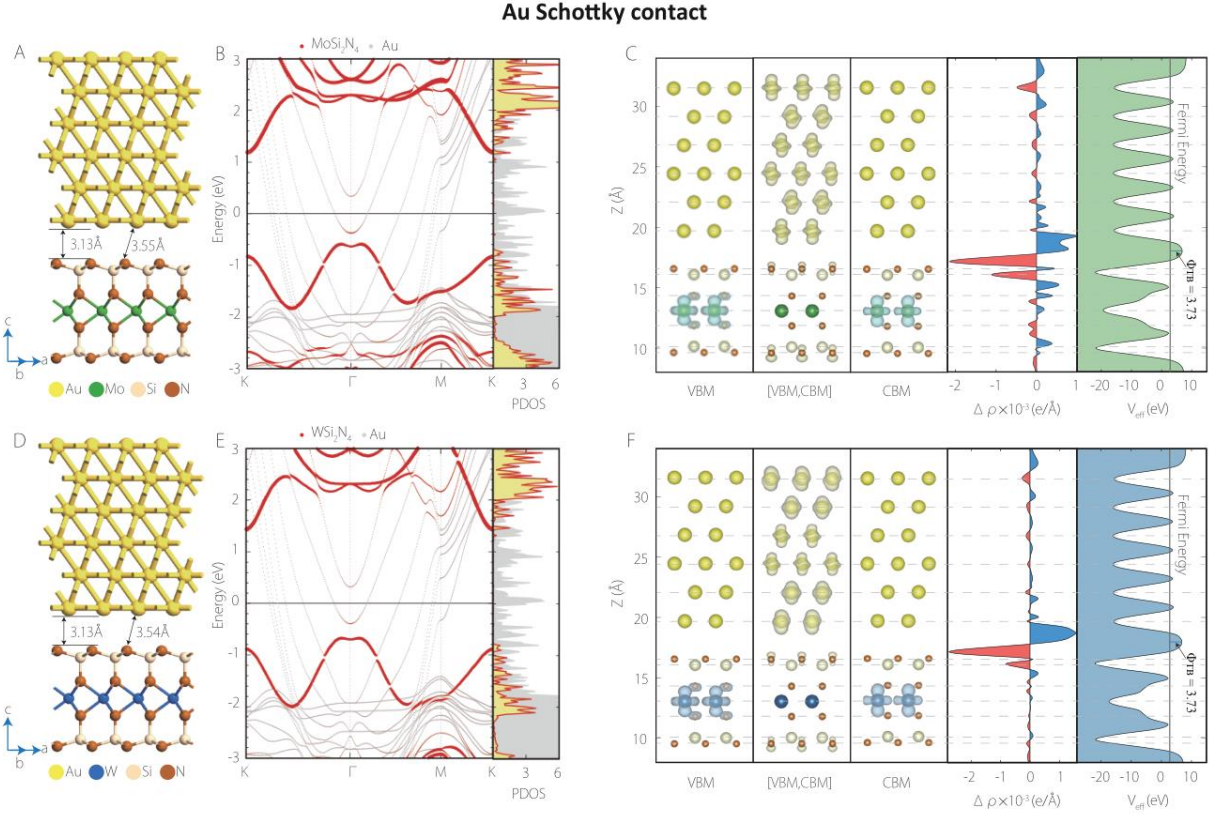

**Figure 2.** Schottky contacts of $Au/MoSi_2N_4$ and $Au/WSi_2N_4$ - Structural and electronic properties of. (A) The contact is composed of 6 layers of Au atoms contacting the $MoSi_2N_4$ monolayer. The interlayer distance is 3.13 Å and the minimum distance between Au and N atoms is 3.55 Å. (B) the projected electronic band structure and the density of states of $Au/MoSi_2N_4$ contact. (C) The panels, from left to right, show the spatial charge density distribution of VBM states, mid-gap states between VBM and CBM, and the CBM states, the differential charge density, and the electrostatic potential profile across the heterostructures. (D), (E) and (F), same as (A), (B) and (C), respectively, for $Au/WSi_2N_4$ contact. The interlayer distance is 3.13 Å and the minimum distance between Au and N atoms is 3.54 Å.

We further calculate the spatial charge density distribution around the VBM and the CBM, as well as the states between the VBM and the CBM. Intriguingly, the semiconducting electronic states are embedded deeply within the inner Mo-N (Figures 2C and 3C) and W-N core layer [Figs. 2(f) and 3(f)], while the mid-gap states between the VBM and CBM are mostly located in the metals and a sparsely distributed in the outlying Si-N layers. In addition, the differential charge density ($\Delta\rho$) reveals a significant charge redistribution across the metal/semiconductor contact interface [Figs. 2(c) and 2(f); Fig. 3(c) and 3(f)], signaling the presence of metal/semiconductor interactions. As $\Delta\rho$ across the contact interface is asymmetrical, the formation of interface dipole is expected to be an important factor that modifies the $S$ parameter from the SM limit (see Fig. S3 for the calculated $\Delta\rho$ of other metal contacts).

Importantly, the absence of mid-gap states in the inner core layer and the finite charge transfer at the outlying Si-N layers reveals that the metal/semiconductor interaction affects mostly the outlying S-N layers, predominantly via charge redistribution, without penetrating the semiconducting Mo-N and W-N inner cores. Here the Si-N layer serves as an encapsulating layer and plays a vital role in preserving the semiconducting characteristics of $MoSi_2N_4$ and $WSi_2N_4$ monolayers. To verify the protective effect of S-N layers, we simulate a *close contact type* of Au/$MoSi_2N_4$ and Au/$WSi_2N_4$ heterostructures [56,57] by forcing the interlayer distance to 1.5 Å, compared to the fully relaxed value of about 3.13 Å (Fig. S3). The semiconducting band structures remain well-preserved at the close-contact limit, thus confirming the robustness of the semiconducting states residing in the $MoSi_2N_4$ and $WSi_2N_4$ monolayers and the resilience against mid-gap states formation. Such unusual behavior, uniquely enabled by the *septuple-layered* morphology of $MoSi_2N_4$ and $WSi_2N_4$ monolayer, is not found in other commonly studied 2D semiconductors, such as transition metal dichalcogenides and black phosphorus, in which an external insertion layer is often required to unpin the Fermi level [58,59].

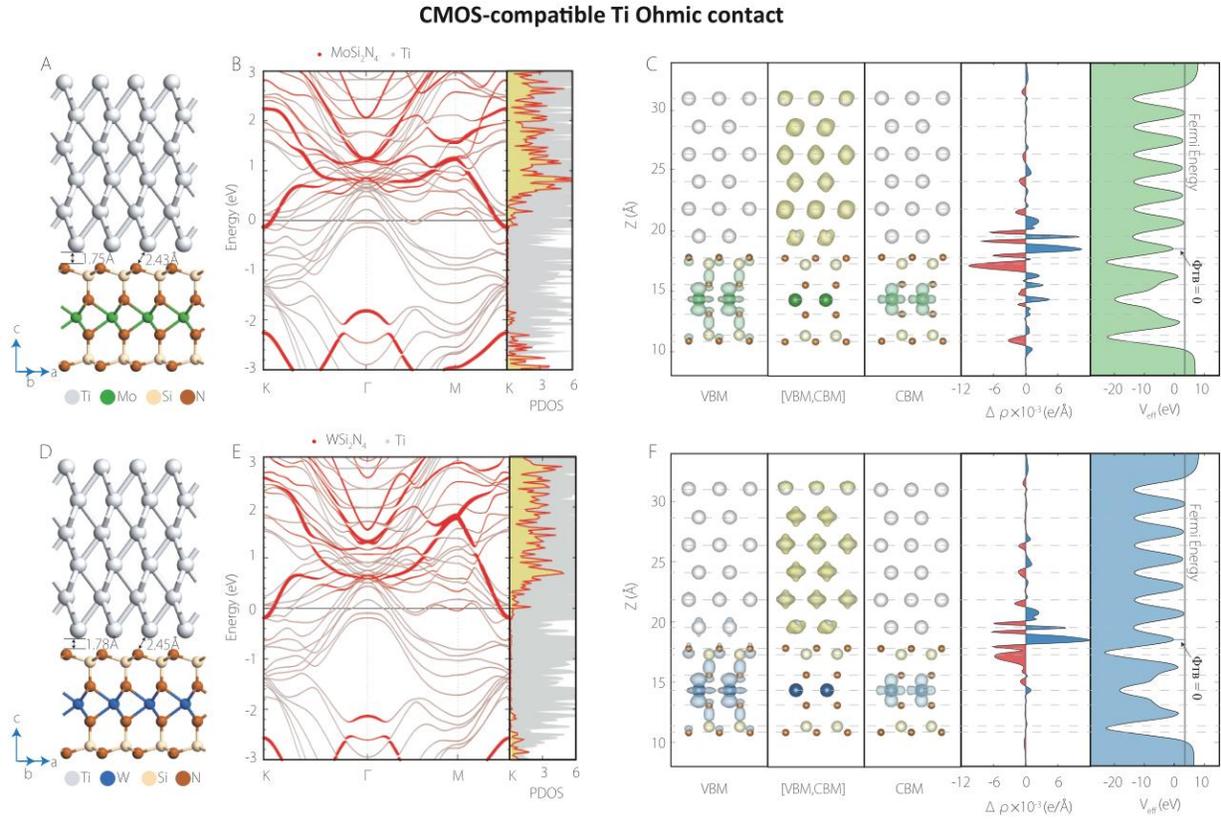

**Figure 3.** Ohmic contacts of Ti/MoSi$_2$N$_4$ and Ti/WSi$_2$N$_4$ Ohmic contacts: Structural and electronic properties. (A) The contact is composed of 6 layers of Ti atoms contacting the MoSi$_2$N$_4$ monolayer. The interlayer distance is 1.75 Å and the minimum distance between Ti and N atoms is 2.43 Å. (B) the projected electronic band structure and the density of states of Ti/MoSi$_2$N$_4$ contact. (C) The panels, from left to right, show the spatial charge density distribution of VBM states, mid-gap states between VBM and CBM, and the CBM states, the differential charge density, and the electrostatic potential profile across the heterostructures. (D), (E) and (F), same as (A), (B) and (C), respectively, for Ti/WSi$_2$N$_4$ contact. The interlayer distance is 1.78 Å and the minimum distance between Ti and N atoms is 2.45 Å.

**Interface potential difference and tunneling potential barrier.** The interface potential difference, $\Delta V = W'_M - W'_S$, where $W'_S$ and $W'_S$ is the work function on the semiconductor and on the metal sides of the contact heterostructures, respectively, lies typically in the range of $\sim 1$ eV (Fig. 4), which is consistent with that of metal-contacted 2D monolayers composed of small atom, such as graphene and hBN [60]. The electronic orbitals of the outermost N atoms of the $MoSi_2N_4$ and $WSi_2N_4$ monolayers are relatively compact and hence more resilient to deformation when compared to that of the contacting metal atoms. By virtue of Pauli exclusion principle, electron density in the interfacial vacuum gap are pushed back towards the metal side, thus leading to a net transfer of electron to metal with $\Delta V > 0$ [61]. The $\Delta V$ is reduced towards zero or a slightly negative value for low work function metals of Sc and In [Figs. 4(a) and 4(b)]. One exception is the graphene contacts. Despite having a moderate work function of 4.91 eV, graphene/$MoSi_2N_4$ and graphene/$WSi_2N_4$ contacts exhibit $\Delta V < 0$. Here, the tug-of-war between the C and N atoms, both are first-row elements with compact electronic orbitals, lead to a slight pushed-back of electrons from the relatively more compact C atom towards the N atom, thus yielding a slightly negative $\Delta V$.

The minimum bond length, $d_{min}$, is defined as the minimum distance between the contacting metal atom and the outermost N atom of $MoSi_2N_4$ and $WSi_2N_4$. In Figs. 4(c) and 4(d), the $\Delta V$ generally decreases with increasing $d_{min}$ due to the weakening of the metal/semiconductor interactions when the metals and the 2D semiconductors are further apart. It should be noted that the Sc, Ti and Ni have relatively shorter $d_{min}$ when compared to other metals, yet the semiconducting electronic band structures remain largely intact and the mid-gap states are strongly suppressed (see Fig. S2). This is again a direct consequence of the outlying Si-N protection layer which prevents the inner core carrying the VBM and CBM electronic states from the perturbation of the contacting metals.

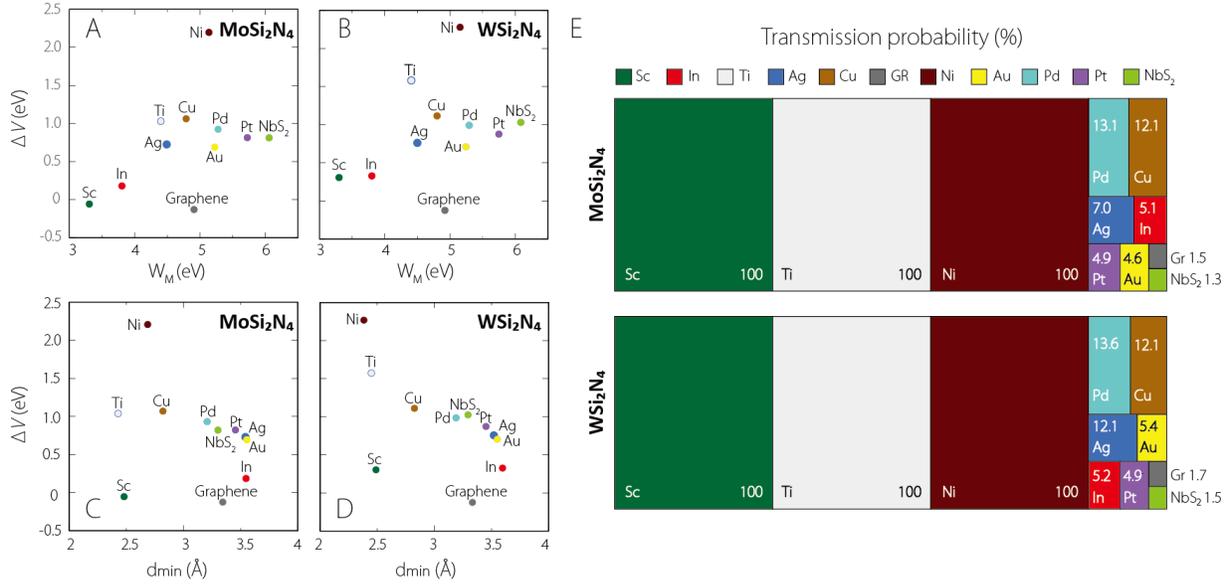

**Figure 4.** Interface potential difference ($\Delta V$) and transmission probability for metal-contacted MoSi$_2$N$_4$ and WSi$_2$N$_4$ monolayers. (A) and (B) shows the isolated metal work function ($W_M$) dependence of $\Delta V$ for MoSi$_2$N$_4$ and WSi$_2$N$_4$, respectively. (C) and (D) shows the minimum bond length ($d_{min}$) dependence of $\Delta V$ for MoSi$_2$N$_4$ and WSi$_2$N$_4$, respectively. (E) Transmission probability of metal/MoSi$_2$N$_4$ and metal/WSi$_2$N$_4$, respectively.

A tunneling potential barrier, $\Phi_{TB}$, can form across the metal/semiconductor gap, which impedes the charge injection efficiency [55]. contact heterostructures. The $\Phi_{TB}$ and the barrier width, $d_{TB}$ is then estimated from effective electrostatic potential across the metal/semiconductor interface (see Fig. S5). For both MoSi$_2$N$_4$ and WSi$_2$N$_4$ monolayers, $\Phi_{TB}$ ranges between 2.8 eV and over 5.1 eV, and the thickness $d_{TB}$ ranges between 1.1 Å and 2 Å (see Tables S1 and S2). The *electron tunneling probability* across the interface is calculated as,

$$T(\Phi_{TB}, d_{TB}) = \exp\left(-\frac{4\pi\sqrt{2m_e\Phi_{TB}}d_{TB}}{\hbar}\right) \quad (3)$$

where $m_e$ is the free electron mass. The $T(\Phi_{TB}, d_{TB})$ calculated via Eq. (3) for various metal contacts are shown in Fig. 4€. For Sc, Ti and Ni, the tunneling barrier is 100% as $\Phi_{TB} = 0$. Particularly for Sc and Ti, both $\Phi_{TB}$ and $\Phi_B$ are zero in both $MoSi_2N_4$ and $WSi_2N_4$ monolayers – an indication of good Ohmic contacts with high charge injection efficiency. In contrast, the VDW type contacts, such as graphene and $NbS_2$ with interlayer distances greater than 3 Å, exhibit very low $T$ of less than 2%, which suggests a low electron transparency in such VDW interfaces.

**Schottky-Mott plot and the $S$ parameter.** As the semiconducting band structures of $MoSi_2N_4$ and $WSi_2N_4$, especially the CBM and VBM states, are well preserved, the electron and hole SBH can be determined from the energy differences between the CBM and the Fermi level ($\varepsilon_F$) of the metal/semiconductor heterostructure, and that between the $\varepsilon_F$ and the VBM via the projected band structures and the PDOS data (see Fig. S2). The Schottky-Mott plot of the metal-contacted $MoSi_2N_4$ and $WSi_2N_4$ is shown in Figs. 5(a)-(d) for both electron-type and hole-type SBH. A linear fit across the 11 metal contacts reveal an remarkable slope parameter of $S_e \approx S_h = 0.69$ for $MoSi_2N_4$ where $S_e$ and $S_h$ denotes electron and hole slope parameters, and $S_e = 0.77$ and $S_h = 0.76$ for $WSi_2N_4$. Although being lower than that of the 2D/2D and 3D/2D VDW-type contacts [38,39], these $S$ values – achieved *intrinsically* without relying on VDW-type contact engineering – are still significantly higher than that of almost all previously reported 2D semiconductors [Fig. 5(e)], such as $MoS_2$ monolayer and bilayer, $WS_2$, InSe, black and blue phosphorene, arsenene, and silicene. Such exceptionally large $S$ values, which spans over a large metal work function range of $\Delta W_M = 2.7$ eV, indicate an enormous flexibility in designing

metal/MoSi$_2$N$_4$ and metal/WSi$_2$N$_4$ heterostructure with a tailor-made SBH as required by the specific device applications.

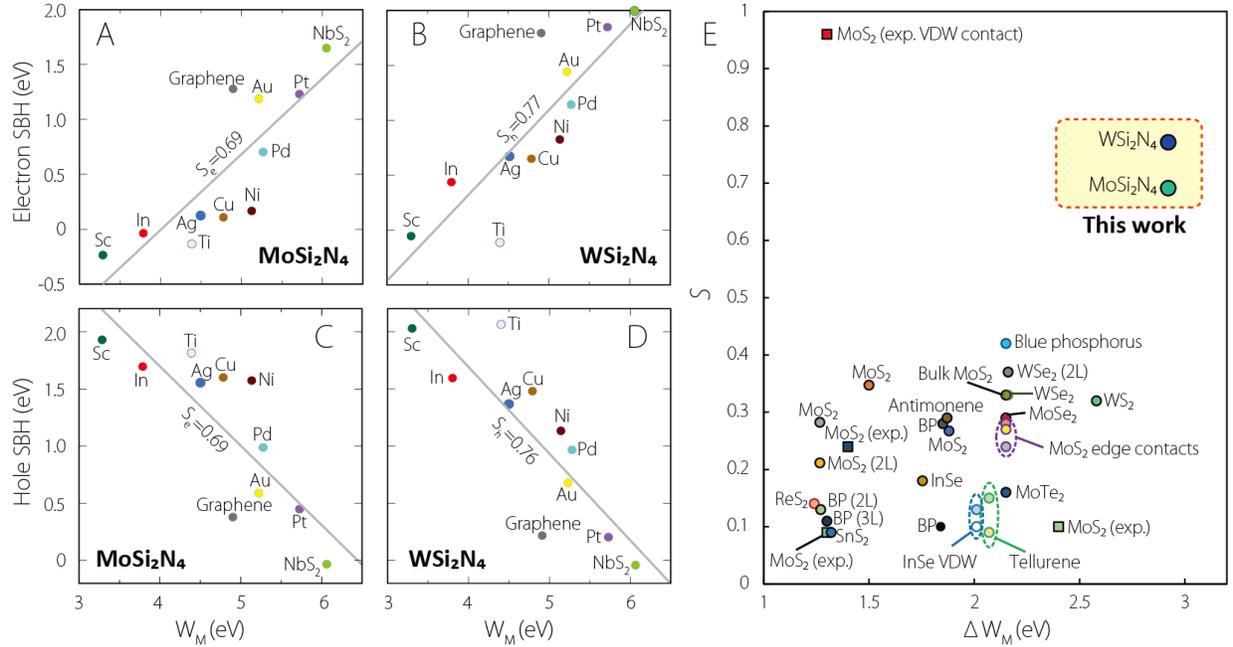

**Figure 5.** Schottky-Mott plot of MoSi$_2$N$_4$ and WSi$_2$N$_4$ under 11 species of metal contacts. (A) and (B) shows the Schottky-Mott plot for the electron SBH of MoSi$_2$N$_4$ and WSi$_2$N$_4$, respectively. (C) and (D) same as (A) and (B), respectively, but for the hole SBH. (E) The slope parameter $S$ versus the metal work function range ($\Delta W_M$) of a large variety of 2D semiconductors, showing exceptionally large S for MoSi$_2$N$_4$ and WSi$_2$N$_4$. Included are the DFT simulations of MoS$_2$ [face contact [19,20,26,27] and edge contact [27]], WS$_2$ [24,27], WSe$_2$ [25], MoSe$_2$ [27], MoTe$_2$ [27], black phosphorus [28–31], blue phosphorus [31], antimonene [32], tellurene [33], InSe [3D metal contact [21] and 2D metal VDW contact [22]], ReS$_2$ [23], and the experimental results of MoS$_2$ [34,35] and SnS$_2$ [36]. The notations '2L' and '3L' denote bilayer and trilayer, respectively.

**Discussion and conclusion.** As interfacial defects and contact imperfections are inevitably presence, the SBH and $S$ values presented in the calculations shall serve as an upper theoretical limit useful for guiding the optimization of device designs and fabrications. Importantly, Ni and Ti contacts, both of which form excellent Ohmic contact with zero interfacial tunneling barrier to $MoSi_2N_4$ and $WSi_2N_4$ monolayers, are CMOS-compatible metals, thus revealing a viable CMOS-compatible contact engineering strategy for the construction of high-performance nanodevices based on $MoSi_2N_4$ and $WSi_2N_4$.

In conclusion, we show that the unique septuple-layered morphology of $MoSi_2N_4$ and $WSi_2N_4$ monolayers offers a fundamentally new mechanism, alternative and complementary to the VDW contact paradigm, for suppressing the FLP effect and for achieving efficient tunable SBH in metal contacts to $MoSi_2N_4$ and $WSi_2N_4$ monolayers. The DFT calculations presented here shall form a harbinger for the study of interfacial contact physics in the expansive family of $MA_2X_4$. Myriads of phenomena, such as the evolution of SBH and FLP with different number of layers [62], the nature of 2D/2D and 2D/3D contacts for the other semiconducting members of the expansive $MA_2X_4$ family, the design of $MA_2X_4$ contact or heterostructures that can facilitate unusual *non-charge* transport, such as spin and valley transport [63], or neuromorphic device operations [15,16], remain to be explored. Future computational and experimental studies of electrical contacts to $MA_2X_4$ shall bring more surprises on the fundamental interface physics and chemistry, as well as the practical design of $MA_2X_4$-based heterostructures and devices [64–66].

## ACKNOWLEDGMENTS

This work is supported by SUTD Start-Up Grant, Singapore Ministry of Education (MOE) Tier 2 Grant (No. 2018-T2-1-007), USA Office of Naval Research Global (ONRG) Grant (No.

N62909-19-1-2047) and SUTD-ZJU IDEA Visiting Professor Grant (SUTD-ZJU (VP) 202001). Q.W. is supported by SUTD PhD Fellowship. S.A.Y acknowledges the support of Singapore MOE AcRF Tier 2 (Grant No. MOE2017-T2-2-108). C.H.L. is supported by Singapore Ministry of Education Academic Research Fund Tier I (WBS No. R-144-000-435-133).

**Figures**

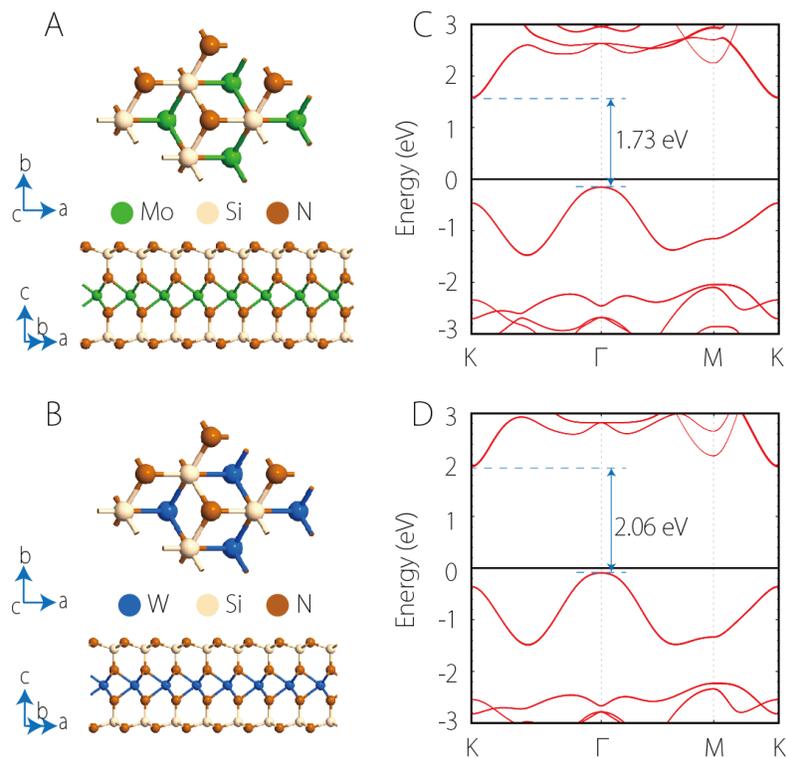

**Figure 1.** Crystal and electronic band structures of isolated $MoSi_2N_4$ and $WSi_2N_4$ monolayers, and a comparison of the slope parameter (*S*) for various 2D semiconductors. Top and side view of (A) $MoSi_2N_4$ monolayer and (B) $WSi_2N_4$ monolayer. Electronic band structures of isolated (C) $MoSi_2N_4$ and (D) $WSi_2N_4$ monolayers.

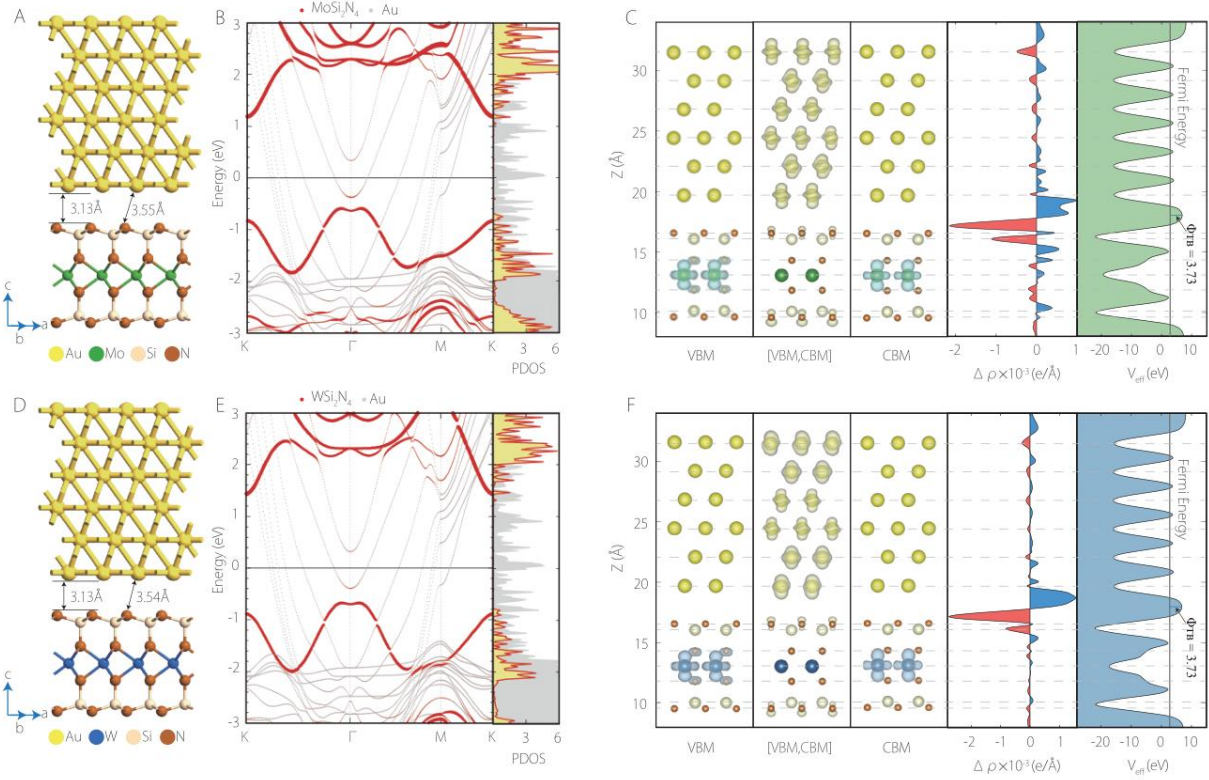

**Figure 2.** Schottky contacts of Au/MoSi$_2$N$_4$ and Au/WSi$_2$N$_4$ - Structural and electronic properties of. (A) The contact is composed of 6 layers of Au atoms contacting the MoSi$_2$N$_4$ monolayer. The interlayer distance is 3.13 Å and the minimum distance between Au and N atoms is 3.55 Å. (B) the projected electronic band structure and the density of states of Au/MoSi$_2$N$_4$ contact. (C) The panels, from left to right, show the spatial charge density distribution of VBM states, mid-gap states between VBM and CBM, and the CBM states, the differential charge density, and the electrostatic potential profile across the heterostructures. (D), (E) and (F), same as (A), (B) and (C), respectively, for Au/WSi$_2$N$_4$ contact. The interlayer distance is 3.13 Å and the minimum distance between Au and N atoms is 3.54 Å.

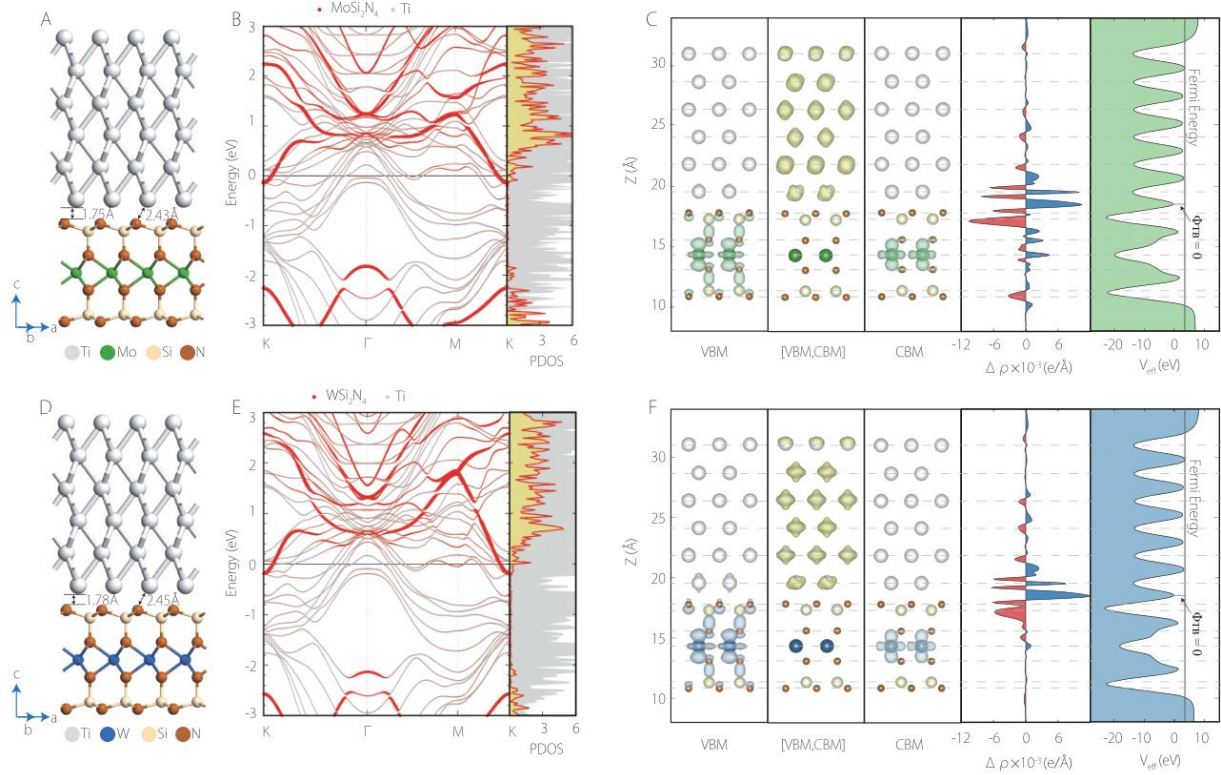

**CMOS-compatible Ti Ohmic contact**

**Figure 3.** Ohmic contacts of Ti/MoSi$_2$N$_4$ and Ti/WSi$_2$N$_4$ Ohmic contacts: Structural and electronic properties. (A) The contact is composed of 6 layers of Ti atoms contacting the MoSi$_2$N$_4$ monolayer. The interlayer distance is 1.75 Å and the minimum distance between Ti and N atoms is 2.43 Å. (B) the projected electronic band structure and the density of states of Ti/MoSi$_2$N$_4$ contact. (C) The panels, from left to right, show the spatial charge density distribution of VBM states, mid-gap states between VBM and CBM, and the CBM states, the differential charge density, and the electrostatic potential profile across the heterostructures. (D), (E) and (F), same as (A), (B) and (C), respectively, for Ti/WSi$_2$N$_4$ contact. The interlayer distance is 1.78 Å and the minimum distance between Ti and N atoms is 2.45 Å.

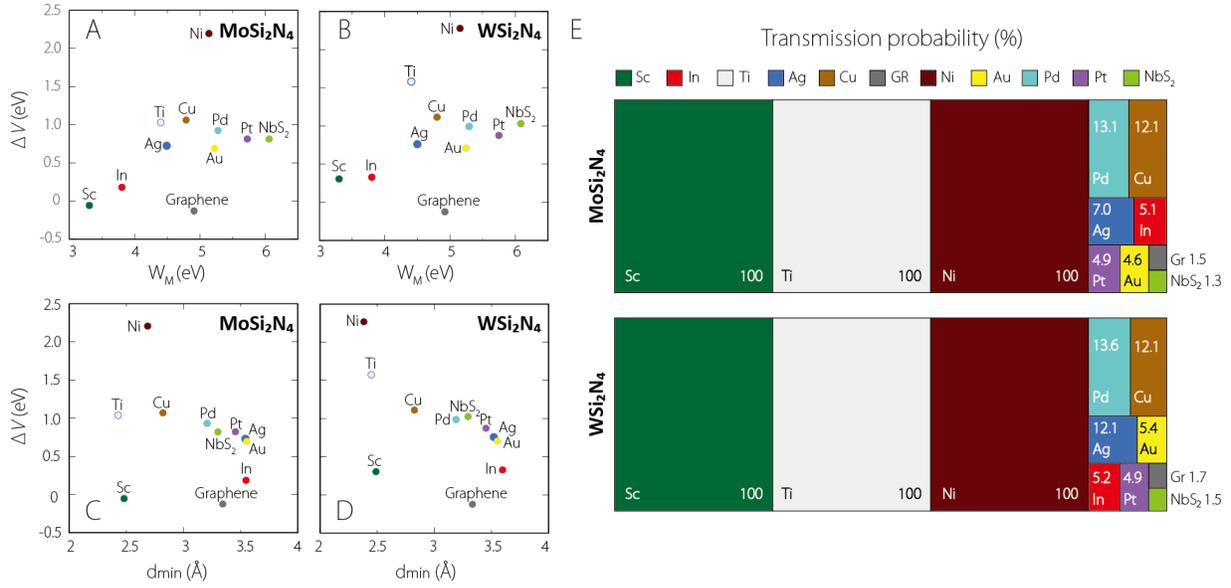

**Figure 4.** Interface potential difference ($\Delta V$) and transmission probability for metal-contacted $MoSi_2N_4$ and $WSi_2N_4$ monolayers. (A) and (B) shows the isolated metal work function ($W_M$) dependence of $\Delta V$ for $MoSi_2N_4$ and $WSi_2N_4$, respectively. (C) and (D) shows the minimum bond length ($d_{min}$) dependence of $\Delta V$ for $MoSi_2N_4$ and $WSi_2N_4$, respectively. (E) Transmission probability of metal/$MoSi_2N_4$ and metal/$WSi_2N_4$, respectively.

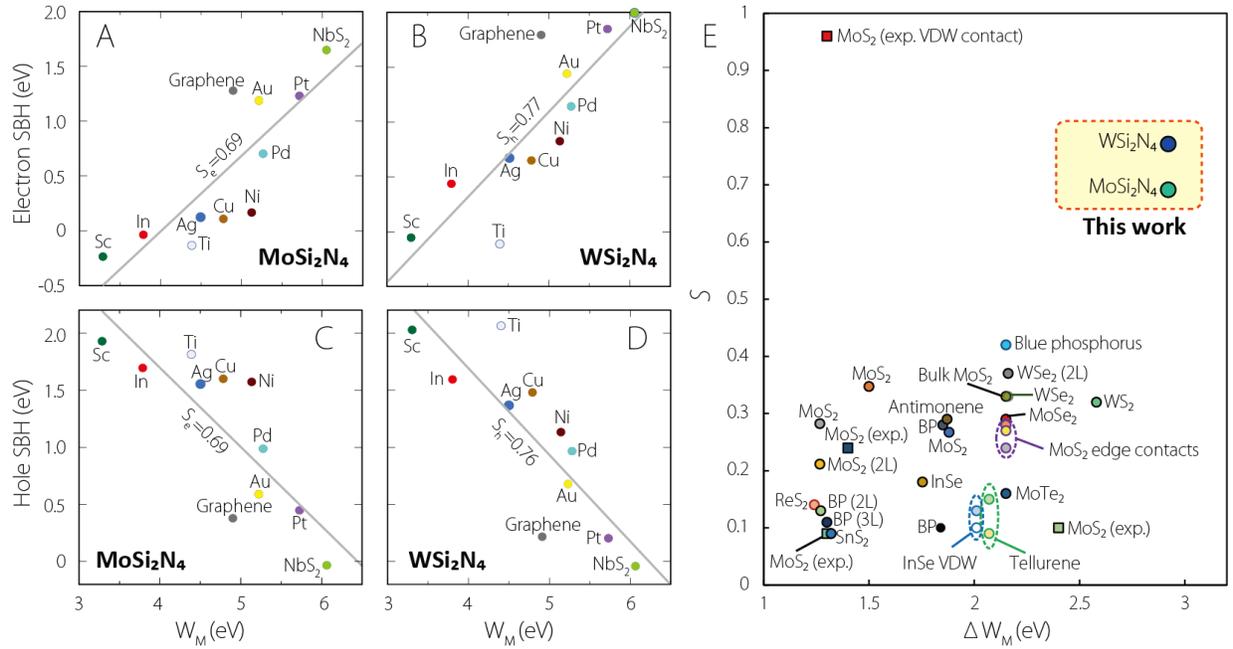

**Figure 5.** Schottky-Mott plot of $MoSi_2N_4$ and $WSi_2N_4$ under 11 species of metal contacts. (A) and (B) shows the Schottky-Mott plot for the electron SBH of $MoSi_2N_4$ and $WSi_2N_4$, respectively. (C) and (D) same as (A) and (B), respectively, but for the hole SBH. (E) The slope parameter $S$ versus the metal work function range ($\Delta W_M$) of a large variety of 2D semiconductors, showing exceptionally large S for $MoSi_2N_4$ and $WSi_2N_4$. Included are the DFT simulations of $MoS_2$ [face contact [19,20,26,27] and edge contact [27]], $WS_2$ [24,27], $WSe_2$ [25], $MoSe_2$ [27], $MoTe_2$ [27], black phosphorus [28–31], blue phosphorus [31], antimonene [32], tellurene [33], InSe [3D metal contact [21] and 2D metal VDW contact [22]], $ReS_2$ [23], and the experimental results of $MoS_2$ [34,35] and $SnS_2$ [36]. The notations '2L' and '3L' denote bilayer and trilayer, respectively.